# Energy-Efficient VM Placement in PON-based Data Center Architectures with Cascaded AWGRs

Mohammed Alharthi
School of Electronic and Electrical Engineering
University of Leeds
Leeds, United Kingdom
elmaalh@leeds.ac.uk

Sanaa H. Mohamed, *Member, IEEE*
School of Electronic and Electrical Engineering
University of Leeds
Leeds, United Kingdom
s.h.h.mohamed@leeds.ac.uk

Barzan Yosuf
School of Electronic and Electrical Engineering
University of Leeds
Leeds, United Kingdom
b.a.yosuf@leeds.ac.uk

Taisir E. H. El-Gorashi
School of Electronic and Electrical Engineering
University of Leeds
Leeds, United Kingdom
t.e.h.elgorashi @leeds.ac.uk

Jaafar M. H. Elmirghani, *Fellow, IEEE*
School of Electronic and Electrical Engineering
University of Leeds
Leeds, United Kingdom
j.m.h.elmirghani@leeds.ac.uk

*Abstract*—Data centers based on Passive Optical Networks (PONs) can offer scalability, low cost and high energy-efficiency. Application in data centers can use Virtual Machines (VMs) to provide efficient utilization of the physical resources. This paper investigates the impact of VM placement on the energy-efficiency in a PON-based data center architecture that utilizes cascaded Arrayed Waveguide Grating Routers (AWGRs). In this paper, we develop a Mixed Integer Linear Programming (MILP) optimization model to optimize the VM placement in the proposed PON-based data center architecture. This optimization aims to minimize the power consumption of the networking and computing by placing the VMs and their demands in the optimum number of resources (i.e., servers and networking devices) in the data center. We first minimize the processing power consumption only and then we minimize the processing and networking power consumption. The results show that a reduction in the networking power consumption by up to 75% is achieved when performing joint minimization of processing and networking power consumption compared to considering the minimization of the processing power consumption only.

## I. INTRODUCTION

Traditional data centre architectures have faced many challenges and limitations for example, low throughput, high latency, limited scalability, management complexity and high cost [1], [2]. Several studies have focused on introducing new energy-efficient designs for the Data Center Networks (DCNs) [3]-[17]. In addition, the growth in the Internet and its applications resulted in an increase in the volume of data that is typically transported over several network domains. This also increased the need for more energy-efficient data centres as the increasing data rates are leading to increase in the number of power hungry devices (i.e. access, aggregation and core electronic switches) within data centers [5], [18]. Several studies developed new architectures that provide solutions for more energy efficient, scalable and reliable data center designs [13], [19], [20].

Among those proposals, different researchers have introduced new designs that replace the power hungry devices by passive components (i.e. couplers and arrayed waveguide grating routers (AWGRs) which offer better power efficiency, lower cost and flexibility in the resource utilization [1]. Using different passive devices, several passive optical network (PON) technologies were proposed for use in modern data centres to provide high performance, energy efficiency, scalability, high capacity and low cost [21].

Virtualization plays an important role in achieving efficient provisioning of physical resource and power saving for data centers. Virtualisation is mostly needed in environments that share resources (i.e. CPU processing and memory) among multiple applications by using a single physical system. According to [22], the computing capacity of data centers is on average 30% utilizated. The 70% of the under utilized resources lead to massive power wastage while keeping the servers running in data centers. Therefore, the under-utilization of the resources of current data centre design is one of the most important challenges. One cause for this is the use of non-energy-efficient resource assignment algorithms where use is not made of resource consolidation. Resource consolidation can be achieved for example via packing VMs in the fewest number of servers and hence switching off a larger number of servers that are now freed.

In this paper, we develop a Mixed Integer Linear Programming (MILP) model to minimize the power consumption by optimizing the VM placement in the PON-based data center architecture we proposed in [23]. This architecture utilizes cascaded AWGRs to increase the scalability of PON-based architectures that use a single layer of AWGRs [12]. We base this study on our previous contributions that utilized MILP optimization models to tackle a range of energy efficiency problems such as processing placement and caching in IoT/Edge networks [24]-[27], greening core and data center networks [28]-[33] by using machine learning, network optimization in data center systems [34]-[37] and the use of network coding to improve the energy efficiency of core networks [38], [39].

The reminder of this paper is organized as follows: Section II describes the system model for VM placement in the cascaded-AWGRs PON data center architecture. Section III explains the proposed MILP model for optimizing the energy-aware routing and VM placement. Section IV provides the optimization results and their discussions. Finally, Section V provides the conclusions and future work.

## II. VM PLACEMENT IN THE PROPOSED DATA CENTER ARCHITECTURE

The connection of the two levels of AWGRs we proposed in [23] are shown in Fig. 1 We proposed using two levels of AWGRs to achieve full and diverse connectivity between the data centre processing cells and between processing cells and OLT switches. This provides flexibility and can enable load balancing in the architecture. In this work, we focus on the optimizing the VM placement in this architecture and its impact on improving the energy efficiency. Our previous work in [23] illustrated more details about the WDM wavelength assignment in the architecture. For inter cell communication, four cells and four OLT switches were considered that were linked through the PON depicted in Fig. 1. For intra cell communications, each cell contains four racks linked by employing a special server (relay server) as shown in Fig. 2. An arrayed waveguide (AWG) multiplexer is used to link the cell to the higher layer AWGRs as shown in Fig1 and in Fig. 2. In addition, Fig. 2 shows that the intra-rack communication is achieved by employing passive polymer optical backplane [40].

For the work in this paper, we assume that the total number of servers in the architecture is 32 where each cell is composed of 8 servers and each rack is composed of two servers. Servers in one cell (i.e., 8 servers in this case) communicate with the remaining servers in the architecture through the special server (relay server). In realistic data center environments, servers have different processing capacities, and hence, we assume that the servers in this work have heterogeneous processing capacities. The special servers require ONUs with tuneable lasers to enable them to direct the traffic from a server in a rack to a server in another rack in the same cell and between servers in different cells by using a dedicated wavelength [20]. The remainder of the servers in this architecture do not have tuneable lasers which reduces costs [20]. For example, if a server needs to communicate with another server in another rack, it communicates first with the special server to request a grant. After that, the special server replies to the server that the request is granted. The server then sets their transceiver to the destination wavelength. TDMA over WDM or any other orthogonal technique can be used to route data between special servers and a given destination server. The communication patterns in this architecture are divided into (i) intra rack communication between servers in the same rack, which is achieved by a passive polymer optical backplane. (ii) Intra cell communication between servers in different racks within a cell, which is achieved by employing special servers with tuneable lasers in the ONUs to connect racks within same cell. (iii) Inter cell communication between servers in different cells. This is achieved through the special servers with tuneable lasers in the ONUs.

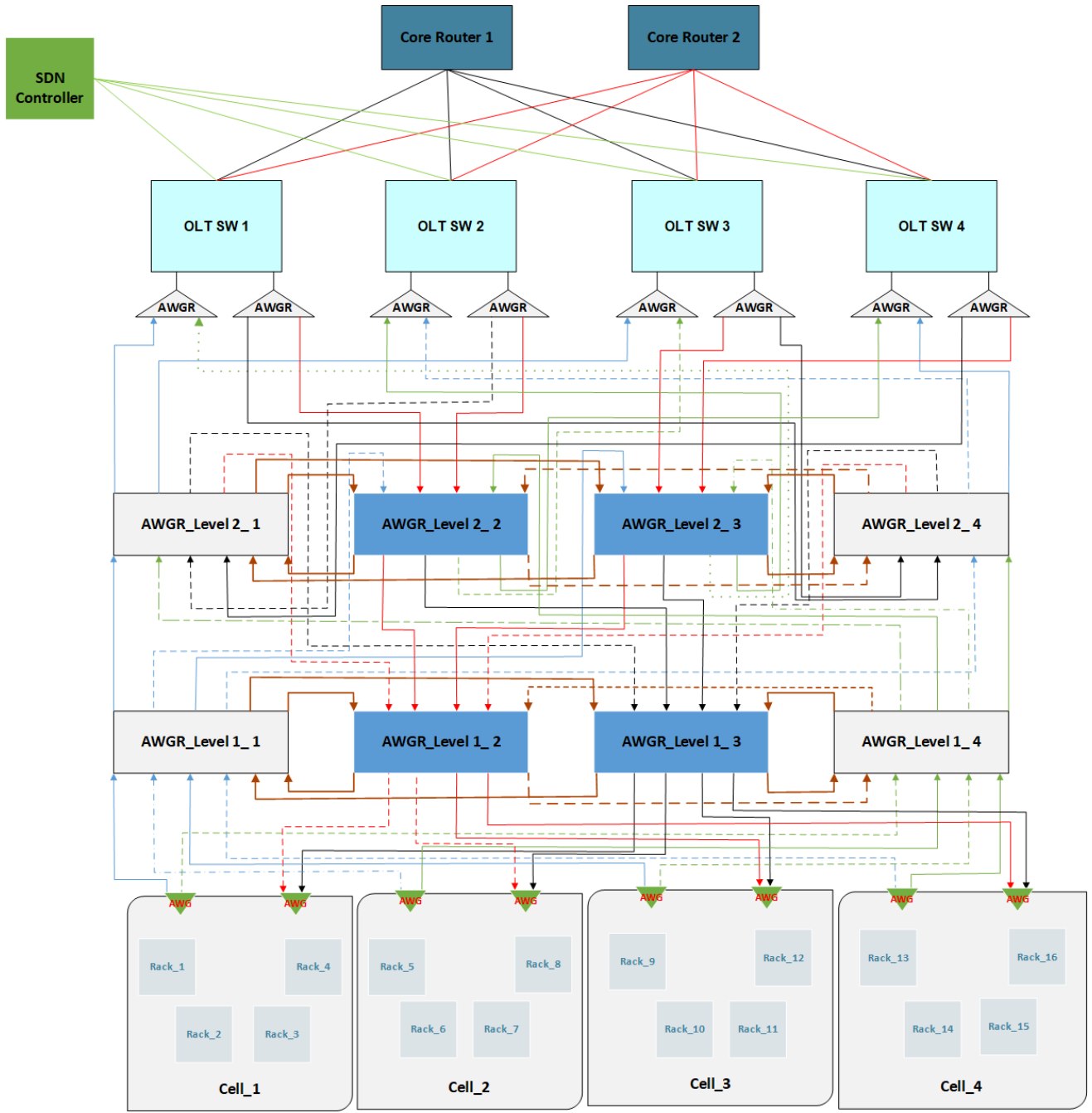

Fig. 1: The Cascaded AWGRs Architecture with four cells.

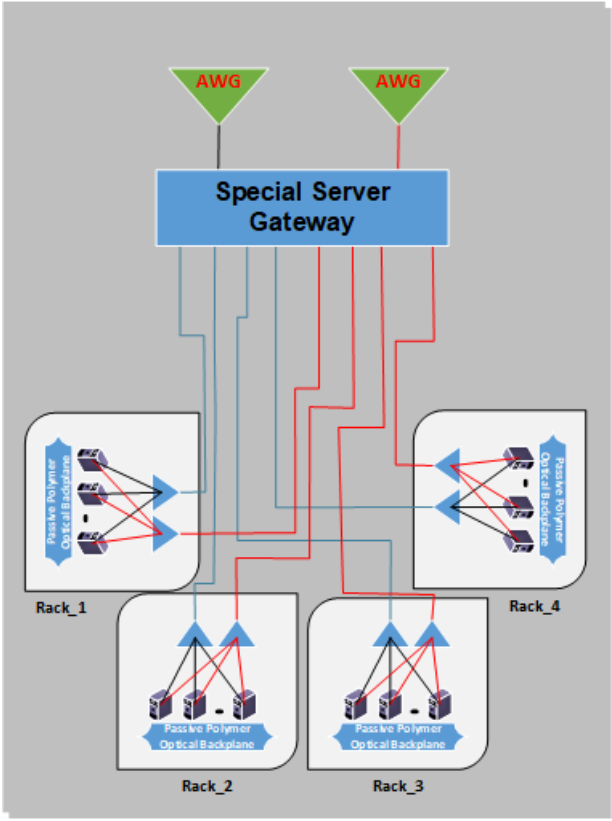

Fig. 2: The intra-cell and intra-rack communication in the cascaded-AWGRs architecture.

### III. MILP MODEL FOR ENERGY-AWARE VM PLACEMENT

In this section, we briefly describe the Mixed Integer Linear Programming (MILP) model we developed for Energy Aware VM placement in the cascaded-AWGRs based architecture. We consider two objectives, the first is to minimize the power consumption of the processing servers and the second objective is to minimize the power consumption of the processing servers, in addition to the networking power consumption of the special servers and their ONUs. The sets, parameters, and variable used in the model are as follows:

Sets and Parameters:

| | |
|---|---|
| $N$ | Set of all nodes in the architecture |
| $N_m$ | Set of neighbouring nodes of node $m$; $m \in N$. |
| $S$ | Set of Servers, where $S \subset N$. |
| $SR$ | Set of special servers, where $SR \subset N$. |
| $TD_{vo}$ | Traffic demand between VMs $v$ and o, where $v, o \in V$. |

Variables:

| | |
|---|---|
| $TR_{sd}$ | Traffic demand between server $s$ and server $d$ resulting from aggregating the traffic of all VMs placed in server $s$, where $(s, d) \in S$. |
| $\beta_{mn}^{sd}$ | Traffic demand between server $(s,d), s, d \in S$, traversing through physical link $(m, n)$, in the architecture, where $m \in N$ and $n \in N_m$. |
| $TTR_s$ | Total traffic forwarded by server s, where $s \in S$. |
| $\vartheta_s$ | The power consumption of server $s, s \in S$. |
| $\mu_r$ | The power consumption of a special server (a relay server) $r, r \in SR$. |
| $\alpha_r$ | Power consumption of ONUs connected to the special server $r, r \in SR$. |
| $SQR_{vs}$ | Is a variable that is equal to 1 ($SQR_{vs} = 1$) if request $v; v \in V$, is served by server $s; s \in S$, and otherwise is equal to zero. |
| $ADRT_{vosd}$ | Is a variable that is the result of ANDing of $SQR_{vs}$ and $SQR_{od}$ and is equal to 1 ($ADRT_{vosd} = 1$) if VMs $v$ and o; $v, o \in V$, are assigned to the different servers $s$ and $k$; $s, k \in S$, and otherwise is equal to zero. |

The power consumption of each processing server is composed of the idle power consumption of this server if it is activated, the proportional power that increases linearly with the VM workload, and the power consumption of the transceiver.

Similarly, the power consumption of special servers is composed of the idle power consumption of the special server if it is activated, the proportional power consumption that is a function of the total traffic they forward in addition to the power consumption of the attached ONU.

The MILP model optimizes VM allocation to the servers in the architecture under one of the following two objectives

*Minimize:*

1. The total processing servers and special servers power consumption (PP & NP):

$$PP \& NP = \sum_{s \in S} \vartheta_s + \sum_{r \in SR} \mu_r + \sum_{r \in SR} \alpha_r \quad (1)$$

2. The processing servers power consumption (PP):

$$PP = \sum_{s \in S} \vartheta_s \quad (2)$$

*Subject to the following constraints:*

1. Traffic and routing constraints:

Constraint (3) is used to calculate the total traffic between server pairs that result from VM processing.

$$TR_{sd} = \sum_{v \in V} \sum_{\substack{o \in V \\ o \neq v}} TD_{vo} \, ADRT_{vosd} \quad (3)$$

$$\forall s, d \in S: s \neq d$$

Constraint (4) expresses the wavelength continuity (i.e., the flow conservation law) which guarantees that a flow entering into a node at a specific wavelength leaves the node at the same wavelength for all the nodes except for the destination and source nodes.

$$\sum_{\substack{n \in N_m \\ m \neq n}} \beta_{mn}^{sd} - \sum_{\substack{n \in N_m \\ m \neq n}} \beta_{nm}^{sd} = \begin{cases} TR_{sd} & m = s \\ -TR_{sd} & m = d \\ 0 & otherwise \end{cases} \quad (4)$$

$$\forall s, d \in S: s \neq d, \forall m \in N$$

Constraint (5) is used to calculate the total traffic transmitted by a server.

$$TTR_s = \sum_{\substack{d \in S \\ s \neq d}} TR_{sd} \quad (5)$$

$$\forall s \in S$$

2. Capacity constraints: We utilized a number of capacity constraints to ensure that the total traffic of a server is within its data rate range, the traffic forwarded by a special server does not exceed the ONU data rate, and that the traffic in each link does not exceed its capacity. In addition, we used capacity constraints to ensure that the demands of the VMs assigned to a server do not exceed its capacity.

## IV. RESULTS AND DISCUSSIONS

Table I illustrates the input parameters used for the model. The model was run while considering different types of VMs requests, processing and memory requirements that were assumed to be uniformly and randomly distributed in this model. A number of VM requests (i.e., 7, 14, and 21) are evaluated.

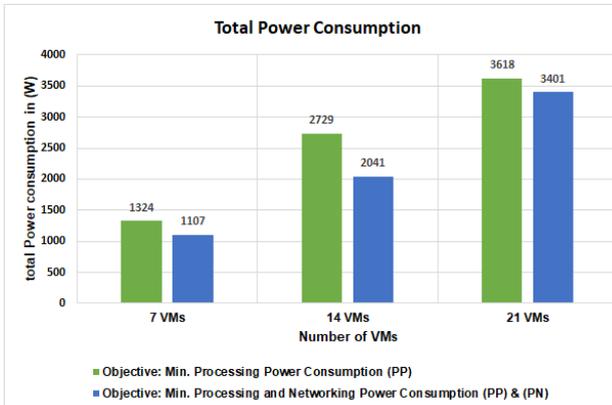

Fig. 3: Total power consumption of different number of VMs under the two objectives.

TABLE I: KEY PARAMETERS FOR THE MILP MODEL.

| Parameter | Value |
|---|---|
| Server's or special server's maximum power consumption. | 301 W [40] |
| Server's or special server's idle power consumption. | 201 W [40] |
| Server's or special server's portion of the processing capacity used for forwarding one request. | 5% |
| Processing capacity of the server. | 6.8 k – 10.8 k (MIPS) |
| Processing capacity of the special server. | 4.30 GHz [41] |
| Memory capacity (RAM) of the server. | 8 - 50 (GB) |
| Data rate of servers. | 1 Gbps |
| Power consumption of ONU. | 2.5 W [42] |
| Data rate of ONU. | 10 Gbps |
| VM's processing request requirements. | random and uniformly distributed, 1.6 k – 10 k (MIPS) |
| VM's memory request requirements. | random and uniformly distributed, 1.7 – 2.7 (GB) |
| Traffic demand between VMs | Random and uniformly distributed, 0.1 – 2.7 (Gbps) |
| Physical link capacity . | 140 Gbps |
| Number of servers Permitted to serve a VM request | 1 |
| Power consumption of server transceiver | 1 W |

Minimizing the power consumption of the processing servers aims to reduce the power consumption of processing by finding the optimal location to place the VMs and this does not take into account the networking power consumption in the architecture. Minimizing the joint power consumption of the processing and networking aims to reduce the overall power consumption of processing and networking devices in the architecture.

The total power consumption results are illustrated in Fig.3 for the different objectives when considering three different number of VMs (7, 14, and 21). The total power consumption under the objective of minimizing the processing and relay servers power consumption (i.e., *PP & NP*) is reduced by 16%, 25% and 6% compared to the objective of minimizing the processing servers power consumption only (i.e., PP) for the evaluated 7, 14 and 21 VMs, respectively.

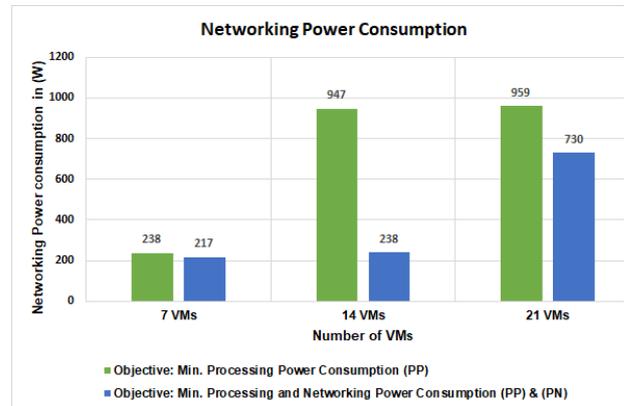

Fig. 4: Networking power consumption of different number of VMs under two objectives

In addition, the networking power consumption is illustrated in Fig. 4 for the different objectives when optimizing the allocation of three different number of VMs (7, 14, 21). The objective of minimizing the power consumption of the processing and special servers (i.e., *PP & NP*) saved the power consumption by 8%, 74% and 23% compared to the objective of minimizing the processing servers power consumption only. This results show that the MILP model tries to consolidate the VMs as much as possible using the least number of servers to avoid the increase in the inter VMs traffic flow between the servers. This in turn, reduces the power consumption of the networking devices (i.e., the special servers and the ONU attached to it).

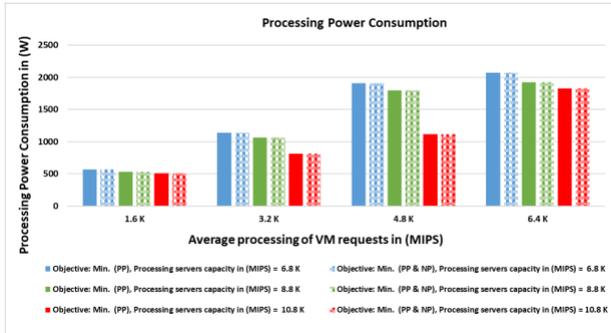

Fig. 5: Processing power consumption of 7 VMs under two objectives with different average processing.

We then examined the impact of increasing the demand of VMs' CPU on the power consumption when considering 7 VMs. In this scenario, the processing capacity of the processing servers was assume to be 6.8 k MIPS (Mega instructions per second), 8.8 k MIPS and 10.8 k MIPS. Figs. 5, 6 and 7 show the power consumption results for each value of the processing capacity for the two considered objectives. As shown in Fig. 5, the power consumption increases as the CPU demands of the VMs increase. This results as a consequence of increases in the number of active physical servers that are needed to accommodate the VMs demand. Moreover, the lowest value of the processing power consumption as shown in Fig. 5 appears clearly at the highest value of servers' capacity as higher capacity leads to serving more VMs in fewer servers. As shown in Fig. 6, with the objective of minimizing the processing and special servers power consumption, a reduction in the power consumption up to 75% was achieved compared to the objective of minimizing the processing power consumption (PP) only.

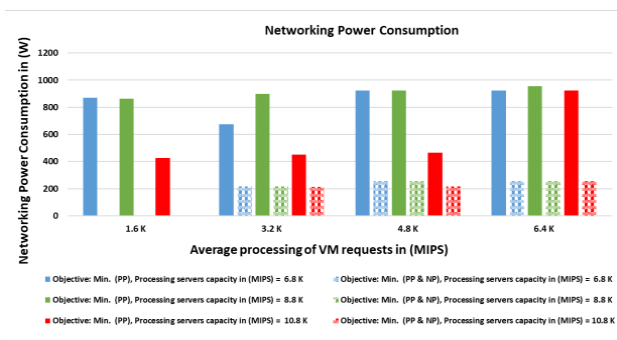

Fig. 6: Networking power consumption of 7 VMs under two objectives with different average processing.

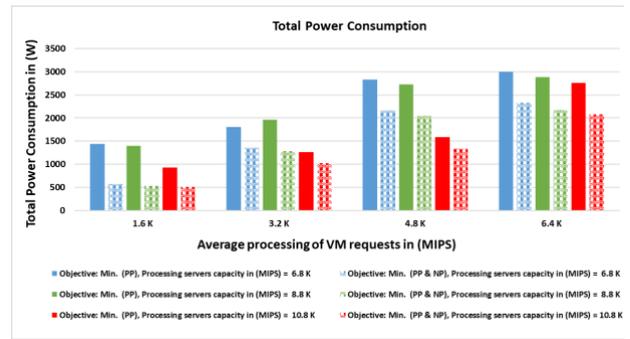

Fig. 7: Total power consumption of 7 VMs under two objectives with different average processing

## V. CONCLUSIONS

This paper introduced a MILP model to minimize the power consumption in a cascaded-AWGRs PON data center architecture by optimizing VMs placement. We considered two objectives. The first is to reduce the power consumption of the processing servers that host VMs and the second is to reduce the power consumption of the processing servers in addition to the power consumption of the special server that is equipped with an ONU for the inter cell communication. The power consumption under the two objectives was tested under different number of VMs (7, 14, and 21). The findings show that the total power consumption under the objective of minimising the power consumption at the processing and networking level (PP & NP) was reduced by 16%, 25% and 6% compared to the objective of minimizing the power consumption at the processing level only (PP) for the evaluated 7, 14 and 21 VMs, respectively. In addition, we considered the impact of increasing the processing requirement for VMs and the capacity of the servers on the total power consumption, processing power consumption and networking power consumption. The results show that the networking power consumption can be reduced by up to 75% when minimizing the total (i.e. processing and networking) power consumption compared to minimizing of the processing power consumption only.

### ACKNOWLEDGMENT

The authors would like to acknowledge funding from the Engineering and Physical Sciences Research Council (EPSRC), INTERNET (EP/H040536/1), STAR (EP/K016873/1) and TOWS (EP/S016570/1) projects. All data are provided in full in the results section of this paper. The first author would like to thank the Ministry of Interior (MOI), Saudi Arabia for funding his PhD scholarship.